# Temperature-Dependent Group Delay of Photonic-Bandgap Hollow-Core Fiber Tuned by Surface-Mode Coupling

YAZHOU WANG,[1,2,†] ZHENGRAN LI,[1,3,†] FEI YU,[1,4,*] MENG WANG, [1] YING HAN,[3] LILI HU,[1,4] AND JONATHAN KNIGHT[5]

[1]*Key Laboratory of Materials for High Power Laser, Shanghai Institute of Optics and Fine Mechanics, Chinese Academy of Sciences, Shanghai 201800, China*
[2]*Center of Materials Science and Optoelectronics Engineering, University of Chinese Academy of Sciences, Beijing 100049, China*
[3] *Key Laboratory for Special Fiber and Fiber Sensor of Hebei Province, School of Information Science and Engineering, Yanshan University, Qinhuangdao 066004, China*
[4]*Hangzhou Institute for Advanced Study, University of Chinese Academy of Sciences, Hangzhou 310024, China*
[5]*Centre for Photonics and Photonic Materials, Department of Physics, University of Bath, Claverton Down, Bath, BA2 7AY, United Kingdom*
[†] *Authors make equal contribution.*
\**yufei@siom.ac.cn*

**Abstract:** Surface modes (SM) are highly spatially localized modes existing at the core-cladding interface of photonic-bandgap hollow-core fiber (PBG-HCF). When coupling with SM, the air modes (AM) in the core would suffer a higher loss despite being spectrally within the cladding photonic bandgap, and would be highly dispersive around the avoided crossing (anti-crossing) wavelength. In this paper, we numerically demonstrate that such avoided crossings can play an important role in the tuning of the temperature dependence of group delay of AM of PBG-HCF. At higher temperatures, both the thermal-optic effect and thermal expansion contribute to the redshift of avoided crossing wavelength, giving rise to a temperature dependence of the AM dispersion. Numerical simulations show that the redshift of avoided crossing can significantly tune the thermal coefficient of delay (TCD) of PBG-HCF from -400 ps/km/K to 400 ps/km/K, approximately -120 ppm/K to 120 ppm/K. In comparison with the known tuning mechanism by thermal-induced redshift of photonic bandgap [*Fokoua et al., Optica 4, 659, 2017*], the tuning of TCD by SM coupling presents a much broader tuning range and higher efficiency. Our finding would provide a new route to design PBG-HCF for propagation time sensitive applications.

## 1. Introduction

In a solid-core silica optical fiber, the thermal-optic effect, thermal expansion and elastic-optic effect (due to the thermal stress) altogether contribute to a strong dependence of modal effective refractive index on the environmental temperature [1,2]. As a result, the fluctuation of ambient temperature inevitably results in thermal-related phase noise on light transmitted through the fiber, which is hazardous for propagation time sensitive applications, e.g. long-distance/distributive time and frequency transfer [3,4], highly accurate time/data synchronization [5] and broadband optical communication networks [6,7].

The thermal coefficient of delay (TCD) is used to characterize the temperature-dependence of group delay of fiber, which is about 37 ps/km/K for a bare SMF28 without fiber coating [8]. The thermal-optic effect of fiber material contributes approximate 95% of TCD in conventional optical fiber, while the other 5% is due to the thermal fiber elongation [6]. By carefully tuning the thermal expansion coefficient of the coating material, the thermal stress can alter the refractive index of the core region through the elastic-optic effect and reduce the TCD of SMF28 to a few ps/km/K, in a limited temperature range from 0 °C to 40 °C [2]. Meanwhile, more efforts have been devoted to the development of time-phase synchronization techniques instead to dissolve the challenges of optical fiber thermal noise [9-11]. It is noted that most of those active compensation methods are applied only for slowly-time-varying thermal phase noises, owing to the limited bandwidth of electronic devices and circuits [12].

The thermal dependence of group delay in standard fibers arises from the intrinsic material thermal response and is impossible to eliminate directly. In contrast, low-loss hollow-core fibers (HCFs) can confine over 99% of light field in the air/vacuum core, and so the impact of fiber material becomes significantly reduced or even negligible [13]. Low-loss HCFs, e.g. PBG-HCF and antiresonant hollow-core fiber (AR-HCF), exhibit low dispersion, low nonlinearity and low material absorption at the same time, when compared to standard fibers.

The low thermal sensitivity of PBG-PCF was firstly experimentally reported by R. Slavik et al in 2015 and its TCD measured as 2 ps/km/K, which is about 1/18 of SMF-28 [8]. Later, G. A. Cranch et al. reported the use of commercial PBG-HCF for thermal phase noise suppression for the first time [14]. In 2017, E. N. Fokoua et al. studied and demonstrated that at the long-wavelength band edge of the transmission window, the thermally induced red shift of the bandgap could balance the thermal expansion of PBG-HCF length and result in a zero or even negative TCD at specific wavelengths. However, this method comes with inevitable leakage loss, which limits the use of a long fiber length in practical applications [6]. In 2019, R. Slavik et al. demonstrated ±50 mrad/m/°C phase temperature sensitivity coefficient of PBG-HCF in the temperature range of -73 °C ~ -69 °C, which is three orders of magnitude lower than that of SMF-28 [15]. At the same time, they found PBG-HCF with the open ends in the air would have the fiber thermal sensitivity further decreased. By controlling the density of the air inside PBG-HCF through pressure, close-to-zero sensitivity (within 0.01 ppm/°C) was observed over 100 nm bandwidth at 113 °C [16]. In 2020, the influence of coating material on the thermal properties of PBG-HCF was studied at temperature in the range of -180 °C ~ 25 °C [17].

The low thermal sensitivity and the tunability of TCD by thermal-induced photonic-bandgap shift makes PBG-HCF of great potential for propagation time sensitive applications. In this paper, we report a new route of tuning the temperature-dependent group delay of PBG-HCF by using the surface mode (SM) coupling. The numerical simulation shows that at different wavelength offset from the avoided crossing where SM coupling appears, the TCD of PBG-HCF can be extensively tuned from -400 ps/km/K to 400 ps/km/K. More importantly, the local loss introduced by SM coupling is moderately lower than near the bandedge of transmission window, which paves its way to the practical use. The birefringence of PBG-HCF introduced by SM coupling is also discussed in the paper.

## 2. Principle

Figure 1(a) shows the dispersion of supermode at the avoided crossing in a typical PBG-HCF when SM couples with air modes (AM) at 0°C and 40°C. At a higher temperature, the redshift of the avoided crossing results in the increase of modal effective refractive index.

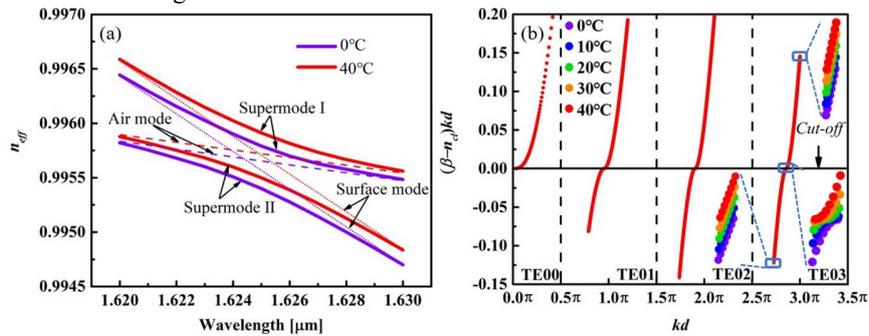

Fig. 1. (a) Simulated supermode dispersions at the avoided crossing of typical PBG-HCF at 0°C and 40°C. The dashed lines show dispersions of independent air (dash) and surface (dot) modes without coupling. (b) The normalized effective refractive indices of $TE_{0m}$ modes in an air-cladding silica glass slab waveguide at different temperatures when thermal optics effects and thermal expansion are considered.

The redshift of avoided crossing at higher temperature is attributed to the redshift of SM dispersion due to the thermal optics effect and thermal expansion. We apply the model of a slab waveguide to approximate the highly localized SM supported by the thin core wall in PBG-HCF. Without losing generality, TE modes are investigated where TM modes are expected to follow a similar trend.

Figure 1(b) plots the dispersion curves of TE modes at different temperature for a silica slab waveguide clad by the air. At a higher temperature, dispersion curves of all TE modes exhibit a similar tendency of redshift. Here we include the thermal optics effect and thermal expansion in the calculation. Details are found in the Appendix.

The redshift of SM dispersion can be characterized by the shift of cutoff wavelength. The cutoff wavelength of $TE_{0m}$ mode is defined by [18],

$$\lambda_m = \frac{2d}{m}\sqrt{n_{co}^2 - n_{cl}^2} \quad (1)$$

Here, $d$ is the thickness of the core wall, $n_{co}$ and $n_{cl}$ are the refractive index. For the core of silica glass, its thermal optical coefficient $\varepsilon = 11\times10^{-6}$ K$^{-1}$ is bigger than thermal expansion coefficient $\alpha = 0.55\times10^{-6}$ K$^{-1}$ by over one order of magnitude [19]. Therefore the thermal optics effect is expected to dominate the redshift phenomenon. Neglecting the thermal induced core wall thickness expansion, the shift of cutoff wavelength is derived from Eq.1 as,

$$\Delta\lambda_m \approx \frac{2d}{m}\frac{n_{co}}{\sqrt{n_{co}^2 - n_{cl}^2}}\varepsilon\Delta T = \alpha_{eff}\Delta T \quad (2)$$

Where, $\Delta T$ is the temperature difference and $\alpha_{eff}$ is defined as the redshift factor of SM with a unit of m/K.

We assume the redshift of avoided crossing equals $\Delta\lambda_m$ and for $\Delta T << T_0$ where $T_0$ is the ambient temperature, the shifted supermode dispersion curve maintains its profile. Then we have,

$$\Delta n_{eff} = n_{eff}(\lambda - \alpha_{eff}\Delta T) - n_{eff}(\lambda) \approx -\alpha_{eff}\frac{dn_{eff}}{d\lambda}\Delta T \quad (3)$$

Naturally we link the dependence of supermode effective refractive index on temperature with the supermode dispersion as follows,

$$\frac{dn_{eff}}{dT} = -\alpha_{eff}\frac{dn_{eff}}{d\lambda} \quad (4)$$

The TCD is defined by [6],

$$TCD = \frac{1}{c_0 L}\frac{dn_g L}{dT} = \frac{1}{c_0}(\frac{dn_g}{dT} + n_g\frac{1}{L}\frac{dL}{dT}) = \frac{1}{c_0}(\frac{dn_g}{dT} + n_g\alpha) \quad (5)$$

Where $L$ is the optical fiber length, $n_g$ is the modal effective group index (GRI), $c_0$ is the speed of light in vacuum. According to Eq.4, the part of TCD attributed by the redshift of avoided cross is derived as,

$$TCD_{ac} = \frac{1}{c_0}\frac{dn_g}{dT} = \frac{1}{c_0}\frac{dn - \lambda\frac{dn}{d\lambda}}{dT} = -\frac{\alpha_{eff}}{c_0}\frac{1}{d\lambda}(n - \lambda\frac{dn}{d\lambda}) = -\frac{\alpha_{eff}}{c_0}\frac{dn_g}{d\lambda} = -\alpha_{eff}D \quad (6)$$

Where $D$ is the dispersion parameter of supermode with a unit of ps/km/nm. It is noted that the thermal optics effect of air is so tiny that its temperature dependence is ignored in the discussion.

Eq.4 and 5 show that the longitudinal change of fiber length by thermal expansion, the redshift factor of SM/avoided crossing and the dispersion of supermode decide the TCD altogether. Near the avoided crossing, the dispersion of supermode usually experience a strong variation, which offers the possibility of tailoring TCD of PBG-HCF in an unprecedented range.

## 3. Simulation and discussion

### 3.1 Simulation models and method

An accurate depiction of tuning of TCD by SM in PBG-HCF is numerically demonstrated by using the finite-element method (FEM). The multi-physics FEM software *COMSOL* is used where the silica glass is used as the host material of fiber and both the thermal expansion and the thermal optics effects are included in the simulation. The effective index of the fundamental-like supermode at the avoided crossing as function of wavelengths are obtained. In the simulation, the ambient temperature is increased from 0°C to 40°C in steps of 10°C. By fitting the group refractive index (GRI) as function of temperature, the derivative of GRI for temperature is obtained. TCD is finally calculated by Eq. 5.

Two kinds of 7-cell and 19-cell PBG-HCFs are modelled as shown in Fig.2. The geometric parameters are summarized in Table I. The refractive indices of silica glass and air are assumed to be 1.45 and 1 at 20°C, respectively. All relevant mechanical and thermal properties are presented in Table II. To introduce SM effectively, the core wall of PBG-HCF is assumed thicker than the strut of cladding [21]. $\Lambda = 4$ and $D/\Lambda = 0.97$ are selected for pitch and core-diameter-to-pitch ration which decides the transmission window centered around 1.55 μm wavelength.

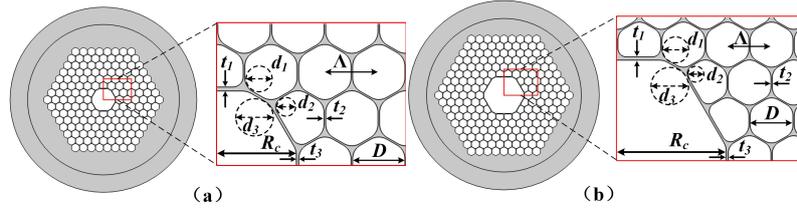

Fig. 2. Models of (a)7-cell and (b)19-cell PBG-HCFs in the simulation. Two models share the same structural parameters except the core diameter $Rc$, $\Lambda$ is the pitch, $d_1$, $d_2$, and $d_3$ are the chamfer diameters of the air hole in the cladding, the inner side of the air hole in the innermost circle, and the core, respectively. $t_1$, $t_3$ and $t_2$ are the wall thickness of the core in the horizontal and vertical directions, and the cladding struct, respectively.

**Table I Structure parameters of the 7-cell and 19-cell HC-PBGF**

|  | $\Lambda$ | $D$ | $d_1$ | $d_2$ | $d_3$ | $t_2$ | $t_1$ | $t_3$ | $R_c$ |
|---|---|---|---|---|---|---|---|---|---|
| 7(19)-cell | 4μm | 0.97$\Lambda$ | 0.6$D$ | 0.4$D$ | 0.7$D$ | $\Lambda$-$D$ | 2.5$t_2$ | 1.5$t_2$ | 1.5 (2.5)$\Lambda$- $t_2$ |

**Table II Material parameters of the silica and air [6]**

|  | Thermal-optical coefficient | Thermal expansion coefficient | Young's modulus | Density | Poisson ratio |
|---|---|---|---|---|---|
| Silica | 11×10$^{-6}$(1/K) | 0.55×10$^{-6}$(1/K) | 72.5(GPa) | 2202(kg/m$^3$) | 0.17 |
| Air | 0 | 1×10$^{-12}$(1/K) | 0(GPa) | 1.29(kg/m$^3$) | 0 |

### 3.2 Simulations and discussion of TCD

In our models, the core wall thicknesses are different in x and y directions, resulting in modal birefringence similar to [21]. Figure 3 and 4 summarize all the simulation results which are categorized by mode polarization.

### 3.2.1 7-cell PBG-HCF

For 7-cell PBG-HCF, one single avoided crossing is identified in the transmission window from 1.5 μm to 1.8 μm, where the coupling of SM with x-polarized AM occurs at a shorter wavelength than the y polarized. The modal effective refractive index and loss at 20 °C are shown in Fig. 3(a-I) and 3(a-II). The simulated increase of local loss at the avoided crossing is found to be of the order of 10$^{-3}$ dB/m only, which is negligible for most practical applications. However, it is noted that tens or hundreds of dB/km losses because of surface mode coupling were experimentally observed [22, 23], which would ultimately limit the applicable fiber length.

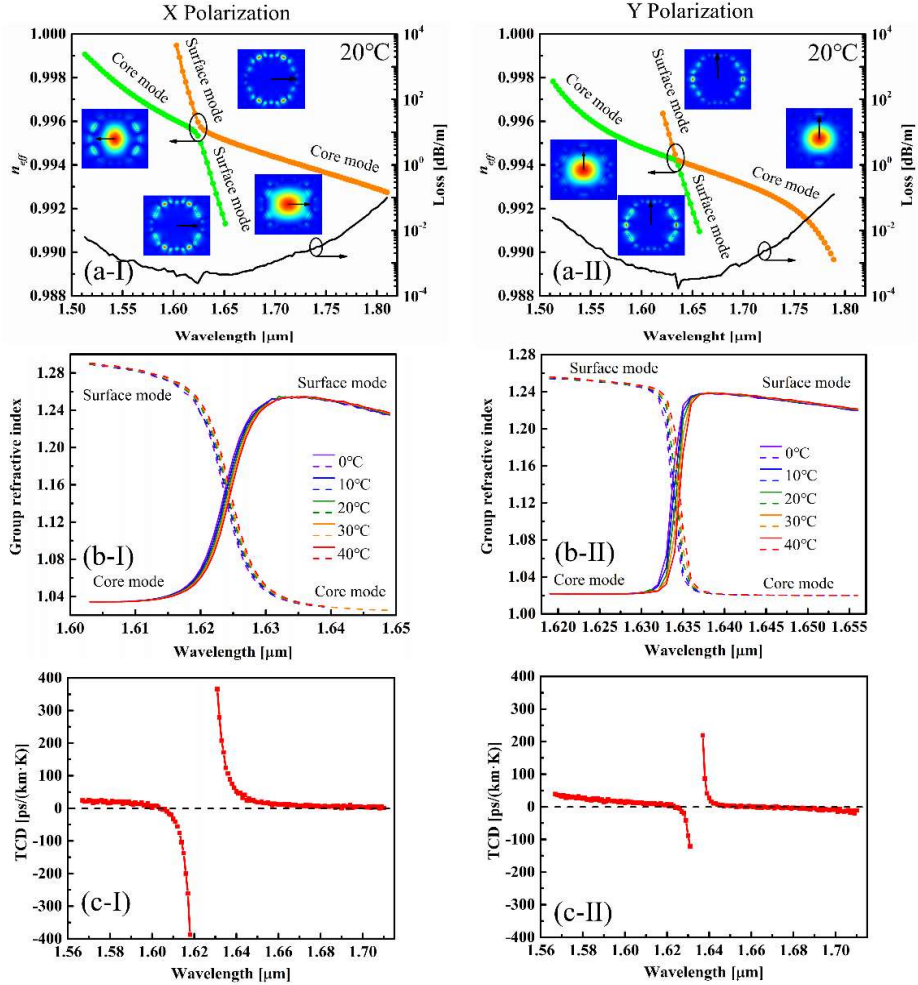

Fig.3 Simulation results of TCD of 7-cell PBG-HCF. Left: x-polarization. Right: y-polarization. (a) The effective refractive index and loss of supermodes near the band gap at 20°C. Rainbow dotted lines: The effective refractive indices; Black solid line: The loss of the core-like supermode. (b) The GRI near the avoided crossing regions at temperature from 0°C to 40°C (c) Calculated TCD of 7-cell PBG-HCF at the avoided crossing.

When the temperature rises from 0 °C to 40 °C, the avoided crossing shifts towards the longer wavelength as analyzed in Section 2. The calculated GRI at the avoided crossing regions for different temperatures are shown in Fig. 3(b-I) and(b-II). The redshift of GRI becomes more significant at wavelengths approaching the center of the avoided crossing resulting in a larger absolute value of TCD. The calculated TCD ranged from -400 ps/km/nm to 400 ps/km/nm which is a much broader range than the tuning by the redshift of photonic bandgap at the band edge as in [6].

Away from the avoided crossing, TCD quickly drops and passes zero. At wavelengths further apart, the modulation of AM dispersion by the coupling of SM is expected to be much weaker and eventually disappears. In this region, the longitudinal thermal expansion starts to dominate TCD of PBG-HCF.

*3.2.2 19-cell PBG-HCF*

TCD evolution near the avoided crossing in the 19-cell PBG-HCF is shown in Fig.4. The birefringence of 19-cell PBG-HCF becomes significant in the vicinity of avoided crossing. For y polarization, avoided crossing I and II are so close that TCD experiences a dramatic change.

As [20] points out, the formation of a bandgap is the consequence of strong coupling of massive cladding modes of PBG-HCF. We argue that the mechanism of SM redshift can be also applied to the redshift of photonic bandgap of PBG-HCF. Near the photonic bandgap edge, the interaction between the AM and specific cladding modes becomes stronger so that the dispersion of AM is modulated and thus contributes to the tuning of TCD. At a higher temperature, the increase of refractive index of cladding material is expected to shift the dispersion curves of cladding modes toward the longer wavelength side, and tune the AM near the band edge in a way as similar to around the avoided crossing. In fact, our TCD curve between two neighboring avoided crossings tuned by SM coupling appears almost the same characteristic as TCD tuned by the redshift of photonic bandgap without SM in [6].

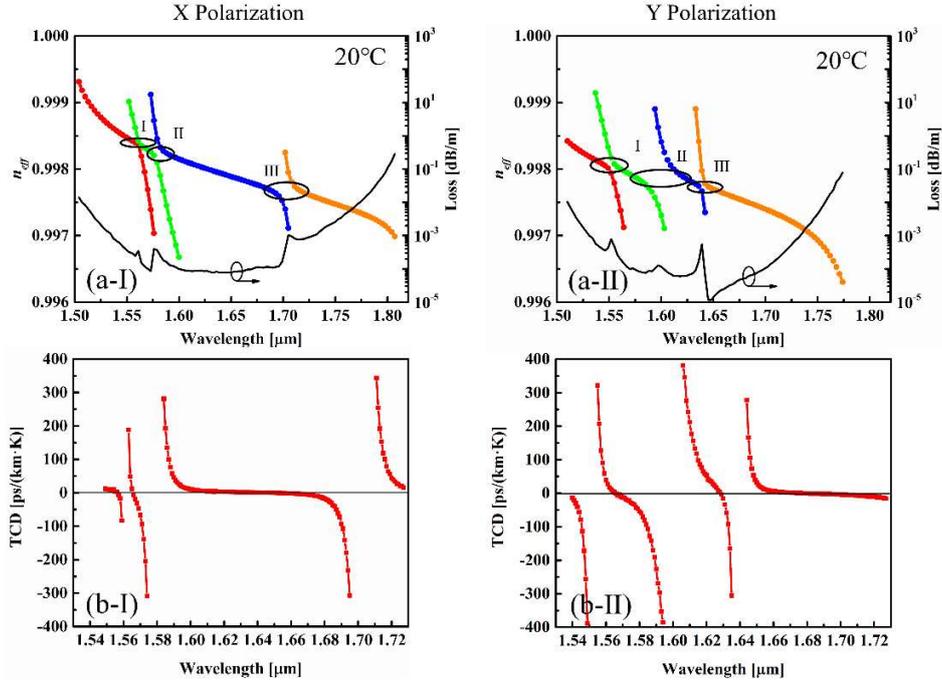

Fig.4 Simulation results of TCD of the 19-cell PBG-HCF. Left: x-polarization. Right: y-polarization. (a) The effective refractive index and loss of supermodes near the band gap at 20°C. Rainbow dotted lines: The effective refractive indices; Black solid line: The loss of the core-like supermode. (b) Calculated TCD of 19-cell PBG-HCF at the avoided crossing.

*3.3 On temperature dependence of birefringence*

Without SM coupling, the birefringence of PBG-HCF is mainly the geometric birefringence determined by the core shape. The thermal induced core expansion in the simulation is so small that the birefringence is regarded as constant. With SM coupling, the birefringence of PBG-HCF is found much enhanced near the avoided crossing because of its strong polarization dependence [21]. Consequently the redshift of avoided crossing is also expected to result in a notable temperature dependence of birefringence.

Similarly we define the thermal coefficient of (phase) birefringence (TCB) to describe temperature dependence of birefringence as,

$$TCB = \frac{dB}{dT} = \frac{d(n_x - n_y)}{dT} \tag{7}$$

As in Eq.3, we have,

$$\Delta B = B(\lambda - \alpha_{eff}\Delta T) - B(\lambda) \approx -\alpha_{eff}\frac{dB}{d\lambda}\Delta T \tag{8}$$

Again we link the dependence of birefringence on temperature with the birefringence dispersion as follows,

$$\frac{dB}{dT} = -\alpha_{eff}\frac{dB}{d\lambda} \tag{9}$$

Figure 5 shows the birefringence, TCB and the derivative of birefringence with wavelength of 7-cell and 19-cell PBG-HCFs. The maximum of birefringence of PBG-HCF reaches $10^{-3}$ at the avoided crossing wavelengths. It is noted that TCB exactly follow the trend of the derivative of birefringence with wavelength but with a negative sign. As the birefringence becomes smooth and flat, the TCB approaches zero as shown.

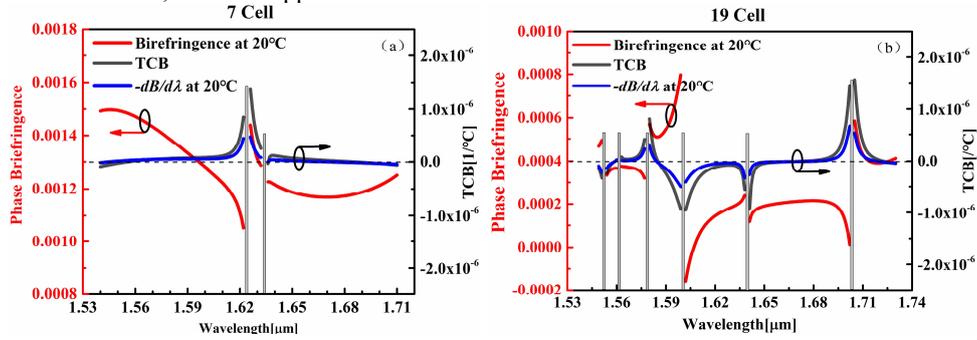

Fig.5 Simulation results of the birefringence, TCB and the derivative of birefringence with wavelength. (a) 7-cell PBG-HCF. (b) 19-cell PBG-HCF. Red solid line: The birefringence of the core-like supermode near the band gap at 20°C. Black solid line: The TCB of the fundamental mode-like supermode. Blue solid line: -dB/dλ of the core-like supermode near the band gap at 20°C. Rectangular shaded are centers of the avoided crossing for x and y polarizations.

## 4. Conclusion

In this paper, we numerically demonstrate that by utilizing the redshift phenomenon of avoided crossing, TCD of PBG-HCF can be flexibly tuned by SM coupling. The tuning range of TCD is greatly extended from -400 ps/km/K to 400 ps/km/K comparing with the reported redshift of photonic bandgap in [6]. The temperature dependence of birefringence of PBG-HCF in the vicinity of SM coupling is also discussed. We point it out that the application of our method for PBG-HCF design for the practical TCD control over a long fiber length will depend on the suppression of increase of local loss by surface mode coupling, which will take more efforts to explore in both theory and experiment in future.

**Funding.** International Science and Technology Cooperation Program (2018YFE0115600); National Natural Science Foundation of China (61935002); Chinese Academy of Sciences (ZDBS-LY- JSC020); National Key R&D Program of China (2020YFB1312802); The Natural Science Foundation of Hebei Province, China (Grant Nos.F2021203002). Fei Yu was supported by the CAS Pioneer Hundred Talents Program.

**Disclosures.** The authors declare no conflicts of interest.

**Appendix**

Figure 1(b) plots the general solution of the eigenvalue equations of $TE_{0m}$ modes in a slab waveguide as approximation of SM of PBG-HCF. The eigenvalue equations of $TE_{0m}$ modes above and below cut-off are written as [20],

$$\begin{cases} W = U\tan U & \text{(Even)} \\ W = -U\cot U & \text{(Odd)} \end{cases}, \text{when } n_{eff} \geq n_{cl}$$

$$\begin{cases} W = iU\tan U & \text{(Even)} \\ W = -iU\cot U & \text{(Odd)} \end{cases}, \text{when } n_{eff} < n_{cl} \qquad (A\text{-}1)$$

Here, $U^2 = (\frac{2\pi}{\lambda}\frac{d}{2})^2(n_{co}^2 - n_{eff}^2)$, $W^2 = (\frac{2\pi}{\lambda}\frac{d}{2})^2(n_{eff}^2 - n_{cl}^2)$, $V^2 = U^2 + W^2$, $n_{co}$ and $n_{cl}$ are the refractive indices of core and cladding and $n_{eff}$ is the effective refractive index of TE$_{0m}$ mode. For a silica glass slab waveguide with the air cladding, $n_{co}$ and $d$ changes because of thermal optics effect and thermal expansion at different temperatures,

$$\begin{aligned} n_{co}(T) &= n_0 + \varepsilon(T - T_0) \\ d(T) &= d_0 + \alpha(T - T_0) \end{aligned} \qquad (A\text{-}2)$$

Here, $n_0$ and $d_0$ are the refractive index and the thickness of the silica slab waveguide at the reference temperature $T_0$. $\varepsilon$ and $\alpha$ are the thermal-optical coefficient and thermal expansion coefficient of silica, respectively. It is noted that thermal effects of air are neglected.